# Ultrafast chiral sensing with an ultraviolet vector beam

**AUDE RODRIGUEZ*** AND **LAURA REGO**

*Instituto de Ciencia de Materiales de Madrid (ICMM-CSIC), C/ Sor Juana Ines de la Cruz 3, 28049 Madrid, Spain*
**aude.rodriguez@icmm.csic.es*

**We present a robust method for imaging molecular chirality by combining ultrafast techniques with vector beams, a type of topological light with inhomogeneous polarization. An infrared vector beam generates high-order harmonics in a sample of randomly oriented chiral molecules, resulting in the emission of an ultraviolet vector beam whose ring-shaped intensity profile reveals the handedness of the chiral molecules. Our calculations predict a highly sensitive discrimination of molecular enantiomers (dissymmetry factor of 186%), opening a new route for studying ultrafast chirality.** 

**Introduction.** The control over the properties of laser light leads to exciting applications across many domains: from everyday technologies (medicine, telecommunications, materials industry, etc.) to cutting-edge research (spectroscopy, imaging, temporal and spatial metrology, quantum entanglement, etc.). One can distinguish between the tailoring of light's microscopic properties (i.e. those defined at a single point in space), such as frequency, polarization, pulse duration or phase, and the control of the macroscopic structure of the beam (i.e. the spatial variation of the light's electric field). Within the latter category, a prominent example is topological light: a type of light that carries a topological charge associated with its macroscopic structure around a field singularity, where the phase and/or polarization is ill-defined. A well-known example of topological light is vortex beams, which carry a phase singularity and whose topological charge is associated with the azimuthal phase variation around it [1]. Alternatively, a new type of topological light has emerged recently: vector beams, which exhibit an inhomogeneous polarization distribution around a polarization singularity. Their topological charge, known as the Poincaré index, is associated with their azimuthal polarization rotation [2]. Vector beams have been shown to enhance super-resolution imaging and nonlinear interactions due to their improved focusing capabilities, and they offer promising applications in optical trapping, communications, or laser manufacturing [3].

Typically, topological light is generated in the infrared (IR) and visible frequency regimes, and it can transfer its properties to higher frequencies through nonlinear interactions [4]. In this manner, it is possible to generate extreme-ultraviolet vector beams via the highly nonlinear process of high-order harmonic generation (HHG) driven in gas [5,6] and solid targets [7,8]. HHG is usually driven by intense ($\sim 10^{13} - 10^{14}$ W/cm$^2$) and very short ($\sim$ fs) IR/visible laser pulses, and results in the emission of extreme-ultraviolet harmonics [9] in the form of attosecond pulses [10,11]. Aside from providing new laser sources, HHG has also opened the door to the study of ultrafast dynamics [12], such as electron motion, probing of matter at nanometric scales due to the short wavelength of the emitted light, and enabling ultrafast spectroscopy techniques [13].

In this theoretical work, we use the ultraviolet (UV) vector beams emitted via HHG as an ultrafast spectroscopic tool to probe a key property of the target: its chirality. Chiral objects are those that differ from their mirror images. In the case of molecules, the two mirror images of a chiral molecule are called enantiomers, and their distinction is highly relevant [14]. Our method is all-optical and based solely on electric-dipole interactions between the laser field and the molecules, leading to high enantio-sensitivity, in line with the so-called electric-dipole revolution [15–17], which includes the use of vortex beams [18]. This contrasts with standard HHG driven in chiral molecules [19–21] or helical dichroism measurements [22–24], as both require the interaction with the magnetic component of the field. Even though vector beams have been employed to drive circular dichroism [25] and optical activity [26] measurements (both also relying on magnetic interactions), as well as to trap [27] or spatially separate enantiomers [28], their use to image chiral handedness via electric-dipole-only (highly sensitive) interactions has not yet been explored. Our elliptically polarized vector beam drives harmonics in randomly oriented chiral molecules, opening a route toward ultrafast enantio-discrimination. The Poincaré index of the driving vector beam is transferred to the harmonics, which therefore preserve the same symmetry as the driving field and form UV vector beams carrying the information of the molecular handedness. We exploit the structure of light to probe a structural property of matter in an isotropic sample, taking advantage of the robustness of topological light.

**Methods.** We propose an all-optical and highly sensitive scheme to distinguish enantiomers using a single driving laser: an elliptically polarized IR vector beam [Fig. 1(a)]. We denote the electric-field components in the transverse plane as $E_\rho$ (radial) and $E_\phi$ (azimuthal) and we omit their spatial dependencies for simplicity. When the IR vector beam interacts with a randomly oriented (isotropic) ensemble of molecules in the form of a thin gas jet, high-order harmonics are generated [Fig. 1(b)]. During this interaction, the component of the induced polarization density, $\vec{P}$, orthogonal to the laser polarization plane, is nonzero only if the molecules are chiral, and exhibits opposite phase for opposite enantiomers [29]. This chiral induced polarization component is the key to achieving an enantio-sensitive measurement. However, under standard focusing conditions, this component is aligned with the propagation and thus does not contribute to the phase-matched harmonic emission.

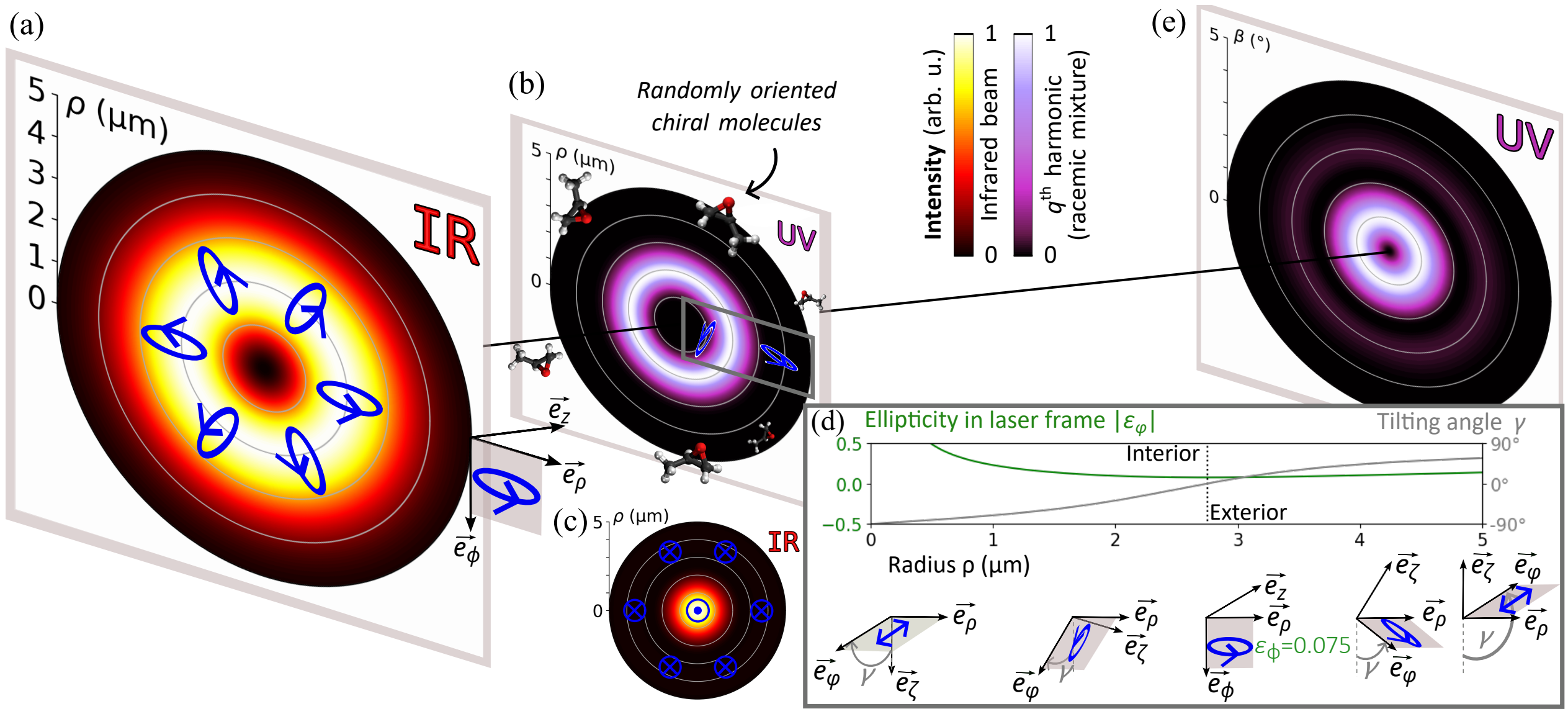

**Fig. 1.** Illustration of ultrafast and topological chiral imaging via HHG. (a) Transverse profile of the IR driving vector beam at the focal plane. Its elliptical polarization is represented by the blue arrows, and the laboratory frame is defined by the unitary vectors $\{\vec{e}_\rho, \vec{e}_\phi, \vec{e}_z\}$. (b) Transverse intensity profile of the $q$th harmonic in the near field, resulting from the interaction between randomly oriented propylene oxide molecules and the driving beam. (c) Longitudinal polarization component of the driving beam appearing upon tight focusing. (d) At the top, the total ellipticity and tilt angle of the driving beam are shown as a function of the radius. The tilted polarization plane is shown at the bottom, where the laser frame unitary vectors are $\{\vec{e}_\rho, \vec{e}_\varphi, \vec{e}_\zeta\}$. The dotted line indicates the change of sign of the tilt angle. (e) Intensity profile of the $q$th harmonic after propagation.

In order to reveal the chiral component, the laser polarization plane must be tilted away from orthogonality to the propagation axis. To achieve this, we tightly focus the vector beam. Indeed, upon tight focusing, a longitudinal field component, $E_Z = \frac{i}{k\rho}\frac{\partial}{\partial\rho}\left(\rho E_\rho\right)$, emerges as part of the IR field [30], as depicted in Fig. 1(c), thereby tilting the local polarization plane of the laser [31]. The tilt of the polarization plane with respect to the propagation direction is backward in the interior of the beam and forward in the exterior part. This is shown in Fig. 1(d), where we have defined the laser frame coordinates $(\rho, \varphi, \zeta)$, which are rotated with respect to the laboratory frame coordinates $(\rho, \phi, z)$ by a local tilt angle $\gamma$ about the radial direction. Note that this tilting also modifies the total ellipticity: the transverse ellipticity, $\varepsilon_\phi = E_\phi/E_\rho$, and the ellipticity in the $\rho z$ plane, $\varepsilon_z = E_z/E_\rho$, combine to give an effective ellipticity in the $\rho\varphi$ plane, $|\varepsilon_\varphi| = (\varepsilon_\phi^2 + \varepsilon_z^2)^{1/2}$. The tilt angle is given by $\gamma = \arcsin(\varepsilon_z/\varepsilon_\varphi)$ and it preserves the axial symmetry of the vector beam. As a result of this tilting, the projections of the chiral and achiral harmonic emissions interfere in the transverse interaction plane, producing an enantio-sensitive near-field intensity pattern. The corresponding far-field UV intensity distribution is displayed in Fig. 1(e). The UV intensity profiles [Figs. 1(b) and 1(e)] correspond to a racemic mixture of left- and right-handed propylene oxide molecules.

To simulate the interaction between the driving beam and the propylene oxide molecules, we first compute the single-molecule response using time-dependent density functional theory (TDDFT) calculations [32] for 200 molecular orientations. We then scale the orientationally averaged microscopic response to the 2D macroscopic beam structure, where the selected harmonic follows a perturbative scaling. Finally, we propagate the emitted radiation to obtain the far-field signal, using Fraunhofer diffraction. The IR vector beam carries a Poincaré index $P = 1$ (defined as the integral of the polarization angle in a closed loop around the singularity, divided by $2\pi$) and a transverse ellipticity of $\varepsilon_\phi = 0.075$. The wavelength is $\lambda_0 = 780$ nm, peak intensity $I = 5.15 \times 10^{13}$ W/cm$^2$, pulse duration 7 fs (full-width half-maximum), and waist $w_0 = 2.75$ μm. Under these conditions, the transverse intensity maximum is reached at $\rho = 1.95$ μm. In Fig. 1(d), the change of sign of the tilt angle $\rho = w_0$ is indicated by a vertical dotted line, and we define the "interior" region where $\gamma < 0$, and the "exterior" region where $\gamma > 0$. The optimal enantio-sensitivity is found for $\lambda_q = \lambda_0/q$ with $q = 5.8$, which we therefore select and refer to as $q$th harmonic. The amplitude at the selected frequency is 3 orders of magnitude lower than the linear response. See Supplement 1 for further details.

**Results.** The induced polarization density in randomly oriented chiral molecules interacting with the laser field is highly nonlinear, and it is composed of three components [see Fig. 2(a)]: two achiral components, $P_\rho$ and $P_\varphi$, lying in the laser polarization plane, and a chiral component, $P_\zeta$, orthogonal to that plane and exhibiting opposite phase for opposite enantiomers. Consequently, the projection of the achiral component $P_\varphi$ onto the laboratory azimuthal direction $\phi$ is identical for both enantiomers, whereas the projection of $P_\zeta$ changes sign, resulting in an enantio-sensitive total polarization, $P_\phi = P_\varphi \cos\gamma + P_\zeta \sin\gamma$. Note that the induced polarization directions determine the dipolar acceleration components, and thus the components of the harmonic emission. The achiral emission generates odd-order harmonics whereas the chiral contribution produces even-order harmonics. Thus, spectral overlap between both channels is required for interference, and is achieved by employing short (i.e., broadband) laser pulses. Under these conditions, the $\phi$-polarized component of the emitted harmonics is enantio-sensitive. In addition, the interior [Fig. 2(a)] and exterior [Fig. 2(b)] regions of the beam contribute differently, since the chiral component of the induced polarization changes sign across the radius. Therefore, the relative phase between the achiral and chiral contributions flips at $\rho = w_0$, leading to a radial modulation of the harmonic emission that depends on the molecular handedness. Figure 2(c) depicts the intensity and phase of the emitted $q$th harmonic order in the laser frame (i.e., the tilted frame). In this frame, only the phase of the $\zeta$ contribution is enantio-sensitive. In the laboratory frame, however, an interference emerges: the achiral and chiral components interfere, giving rise to enantio-sensitive intensity modulations, as shown in Fig. 2(d). A strong difference in peak intensity can be observed, together with an enantio-sensitive spatial variation of the intensity and phase. Both enantiomers present a ring structure, but the emission is stronger for the L-handed enantiomer, while it is suppressed for the R-handed enantiomer, see Figs. 2(e) and 2(f).

Upon propagation to the detection plane, the emission from each enantiomer acquires secondary rings [Figs. 3(a) and 3(b)]. The difference between the intensity profiles of the azimuthal component of the two enantiomers is depicted as a function of the radius in Fig. 3(c). Note that both emissions have a smaller radius than the IR driving beam (red line), such that they can be spatially filtered out. Note as well that the Poincaré index ($P = 1$) is conserved throughout the propagation thanks to the axial symmetry of the vector beam, as can be deduced from the $2\pi$ polarization rotation in Fig. 3(d). The differences in ellipticity reflect that the azimuthal component's phase is also enantio-sensitive. In addition, in a sample containing both enantiomers, the intensity profile allows the determination of their concentrations, as shown in Fig. 3(e), obtained by the coherent addition of the enantiomers' emissions.

Our scheme provides control over the enantio-sensitivity via the laser parameters. Figure 4 shows the far-field intensity and dissymmetry factor, $\mathrm{DF} = 2(I_\alpha^L - I_\alpha^R)/(I_\alpha^L + I_\alpha^R)$, as functions of the transverse ellipticity of the driving field, the selected harmonic order, and the beam waist. First, the transverse ellipticity plays a complex role: it increases both the $E_\varphi$ (achiral) and $E_\zeta$ (chiral) contributions, but reduces the tilt angle, thereby keeping the chiral component largely aligned with the propagation direction. As a result, we chose an ellipticity of 0.075 as a compromise between stronger signal and a significant difference in intensity between the enantiomers. It is also worth noting that changing the sign of $\varepsilon_\phi$ exchanges the results between the enantiomers. Which of the enantiomers exhibits a stronger emission is therefore under our control. Second, the selected harmonic frequency represents a compromise between a large dissymmetry factor and a strong harmonic signal (the full HHG spectrum is shown in Supplement 1). Nevertheless, nearby frequencies still yield pronounced enantio-sensitive emissions. Third, tighter focusing (i.e., smaller waist $w_0$) enhances the longitudinal field component, leading to larger tilt angles; hence our choice of small waist values. However, enantio-sensitive intensity profiles can be obtained over a broad range of waist parameters. Overall, the dissymmetry factor demonstrates the robustness of the results with respect to variations in the driving field parameters and selected harmonic frequency, as well as the degree of control over the resulting enantio-sensitivity. In addition, the waist or ellipticity can be used to balance the projections of the chiral and achiral components if a different molecular species is used.

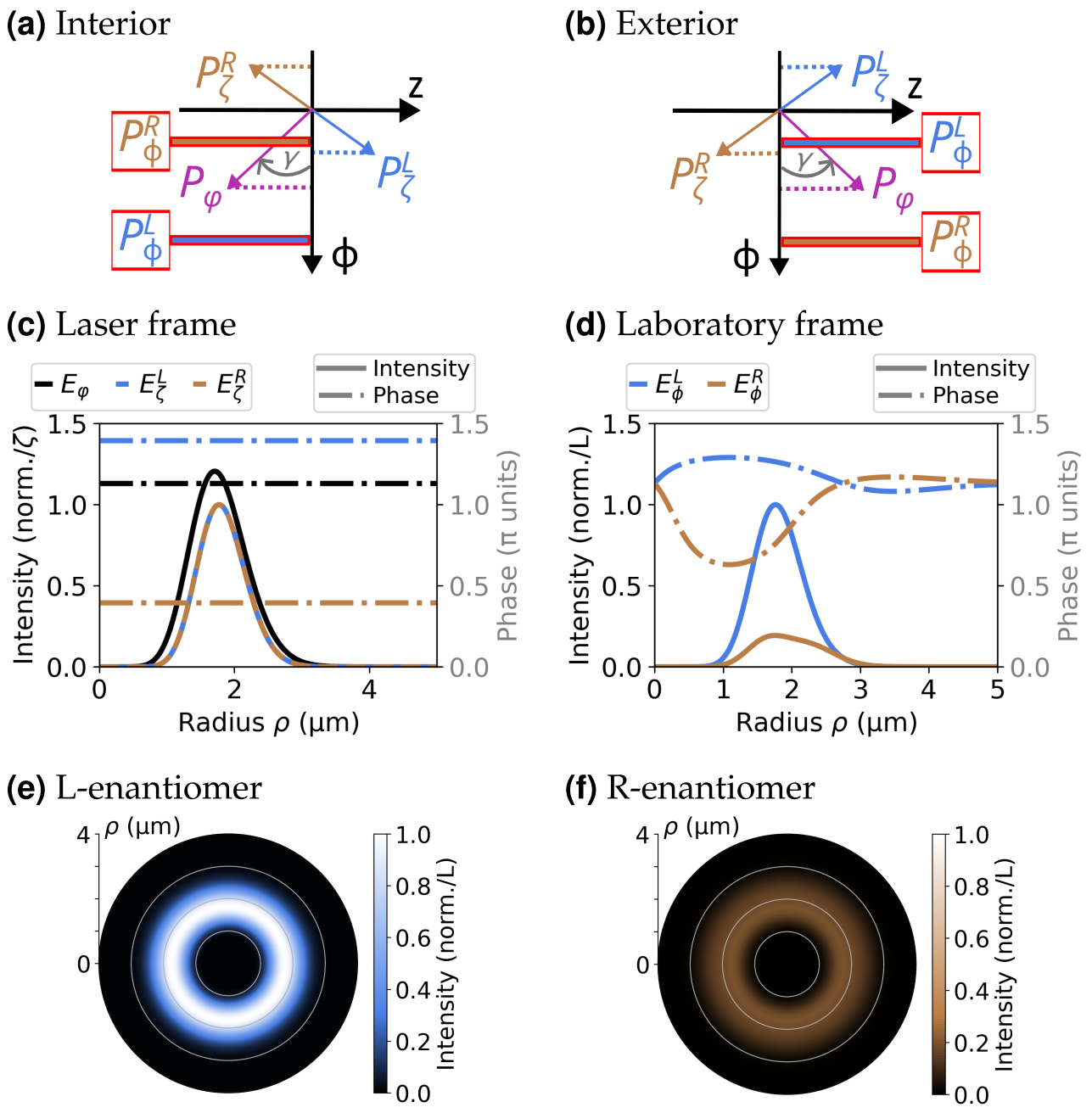

**Fig. 2.** Schematic representation of the tilting of the polarization in (a) the interior and (b) the exterior of the beam, for the L- and R-handed enantiomers (in blue and brown, respectively). The red squares are $P_\phi$ for the two enantiomers. (c)–(f) Near field of the $q$th harmonic. (c) Intensity (solid lines) and phase (dotted lines) of the $E_\phi$ (black, achiral) and $E_\zeta$ (blue for the L-enantiomer, brown for the R-enantiomer) components as a function of the radius in the laser frame. (d) Same as (c) but in the laboratory frame: projections over the $\phi$ axis for the L- (blue) and R- (brown) enantiomers. (e), (f) Transverse intensity profile of $E_\phi$ in the laboratory frame emitted by the (e) L- and (f) R-enantiomers. All intensities have been normalized with respect to the strongest component.

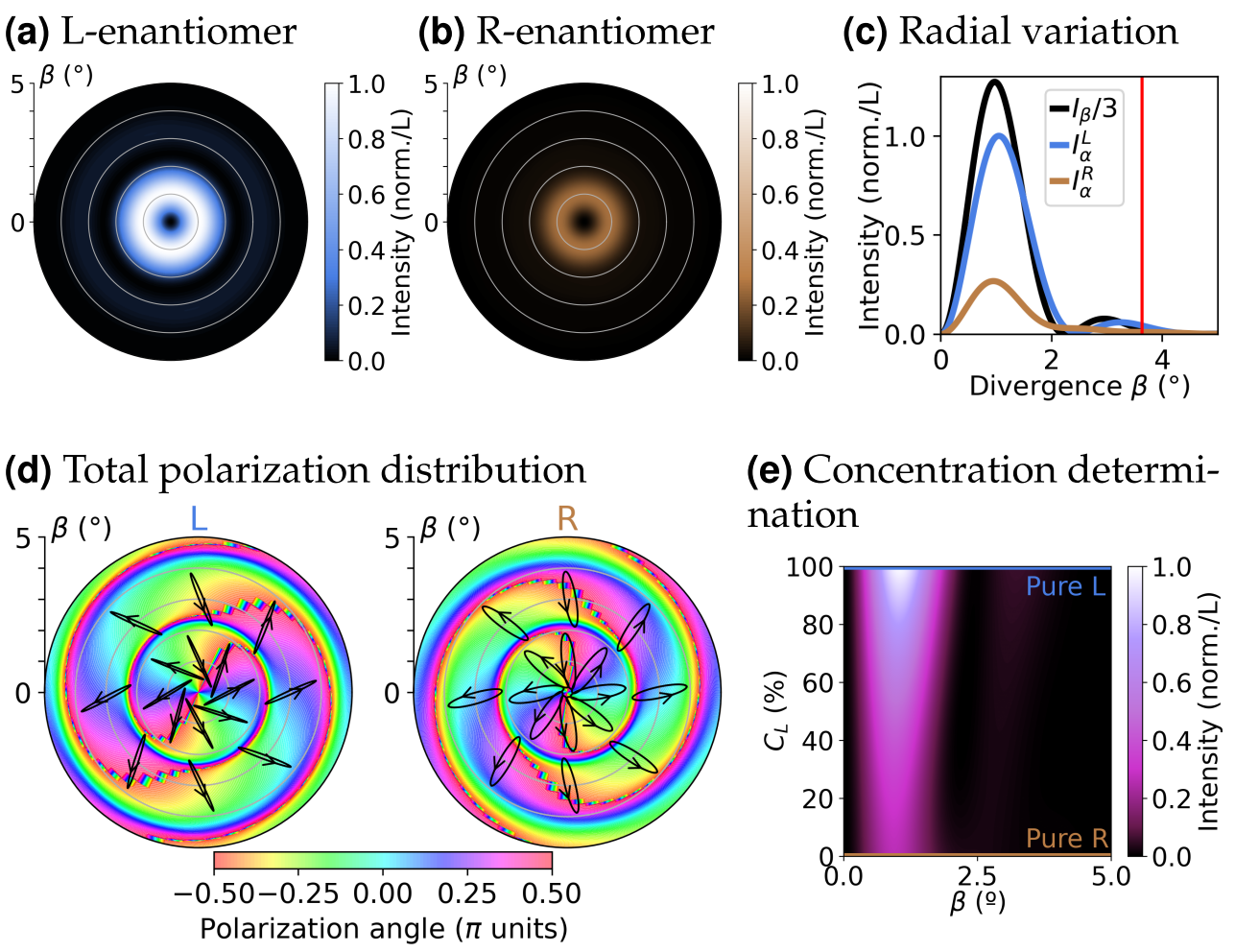

**Fig. 3.** Intensity of the azimuthal component of the $q$th harmonic after propagating to the far-field plane from the (a) L- and (b) R-enantiomers. (c) Radial variation of the intensity profiles of the radial $\beta$ and azimuthal $\alpha$ components. (d) Total polarization distribution of the $q$th harmonic emitted by each enantiomer. (e) Intensity profile of the azimuthal component for different concentrations of the L-enantiomer ($C_L$).

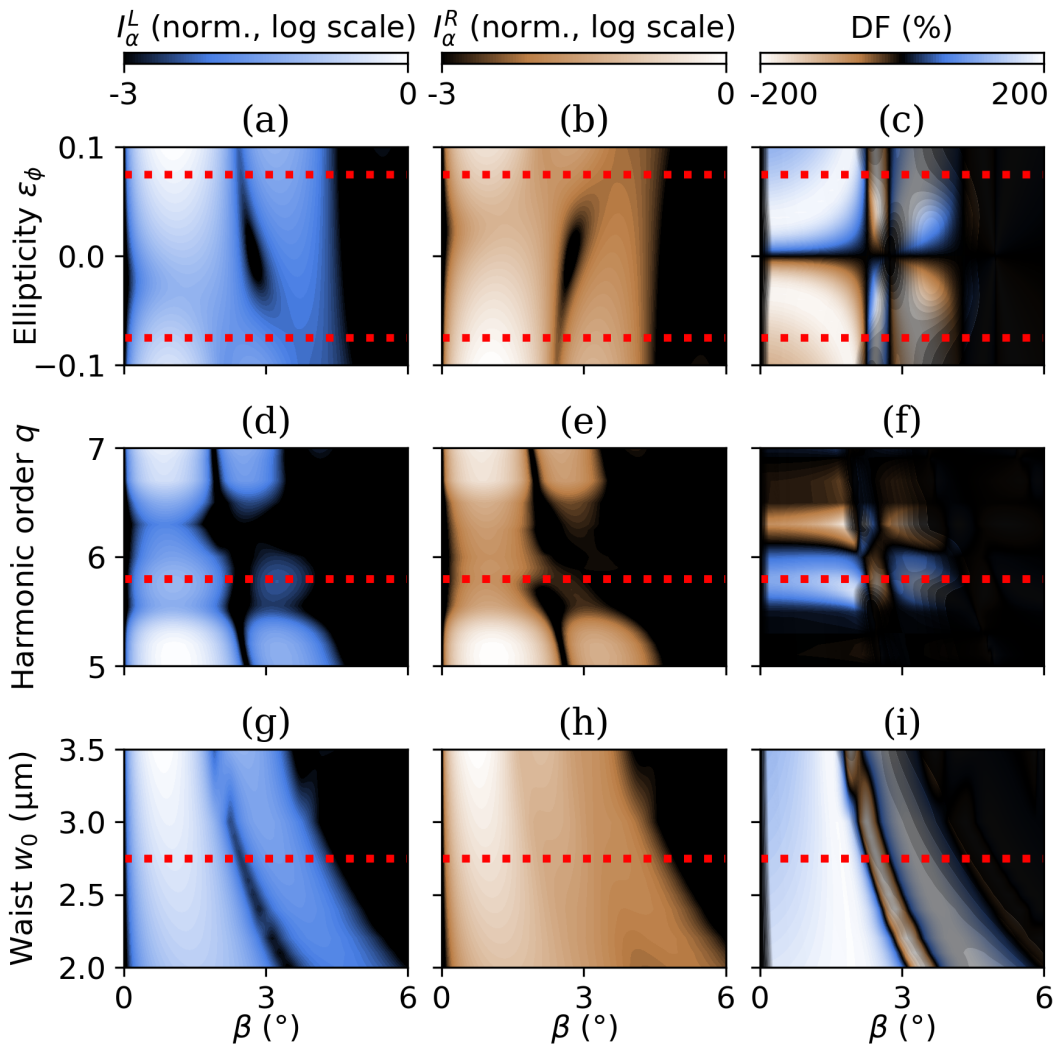

**Fig. 4.** Control over the enantio-sensitivity. (a)–(c) Influence of the transverse ellipticity $\varepsilon_\phi$ of the driving beam on the intensity profiles of the L- (a) and R- (b) enantiomers and the dissymmetry factor (c). (d)–(f) Influence of the harmonic order $q$ on the intensity profiles of the L- (d) and R- (e) enantiomers and the dissymmetry factor (f). (g)–(i) Influence of the waist of the beam $w_0$ on the intensity profiles of the L- (g) and R- (h) enantiomers and the dissymmetry factor (i). The dotted red lines indicate the values with the best enantio-sensitivity. The dissymmetry factor is masked to show only regions where the intensity is above 10% of its maximum.

**Conclusion.** In conclusion, we propose an all-optical and ultrafast approach to distinguish molecular enantiomers using ultraviolet vector beams, where we have numerically demonstrated the encoding of the enantio-sensitivity in the intensity profiles of the emitted harmonics. Compared to a Gaussian beam, a vector beam offers several key advantages: a stronger longitudinal component [2], intrinsic robustness to perturbations due to its topological nature [33], and on-axis emission (in contrast to non-collinear configurations [16] or schemes relying on enantio-sensitive propagation directions [31]). In addition, our configuration is robust with respect to changes in the beam waist and the selected harmonic frequency, while providing tunable control via the ellipticity of the driving field, whose sign exchanges the response of left-handed and right-handed enantiomers. Furthermore, the generation of high-order harmonics in the form of attosecond pulses offers a route toward attosecond temporal resolution, opening the possibility of ultrafast chiral measurements with single-color topological light. Altogether, our scheme connects topological light, chirality, and ultrafast dynamics in a unified framework.

**Funding.** Fundación Bancaria Caixa d'Estalvis i Pensions de Barcelona (LCF/BQ/PR24/12050018); Ministerio de Ciencia, Innovación y Universidades (PID2024-163024NA-I00); Fundación Severo Ochoa (CEX2024-001445-S).

**Acknowledgment.** The authors acknowledge the support and fruitful discussions with Dr. David Ayuso. The project leading to these results has received funding from "la Caixa" Foundation (ID 100010434), under the agreement "LCF/BQ/PR24/12050018," from the Spanish Ministry of Science, Innovation and Universities the State Research Agency through the project ref. PID2024-163024NA-I00 (MICIU/AEI/10.13039/501100011033/FEDER, UE), and from the Severo Ochoa Centers of Excellence program through Grant CEX2024-001445-S.

**Disclosures.** The authors declare no conflicts of interest.

**Data availability.** The findings of this study are available upon reasonable request from the authors.

**Supplemental document.** See Supplement 1 for supporting content.

# Ultrafast chiral sensing with an ultraviolet vector beam: supplement

**Aude Rodriguez*** **and Laura Rego**

*Instituto de Ciencia de Materiales de Madrid (ICMM-CSIC), C/ Sor Juana Ines de la Cruz 3, 28049 Madrid, Spain*
*aude.rodriguez@icmm.csic.es

# ULTRAFAST CHIRAL SENSING WITH AN ULTRAVIOLET VECTOR BEAM: SUPPLEMENTAL DOCUMENT

## 1 Introduction

The simulation of the interaction between the infrared (IR) vector beam and the chiral molecules combines two scales: microscopic and macroscopic. The microscopic level consists in the time-dependent response of an ensemble of randomly oriented chiral molecules to the local elliptically-polarized electric field of the laser (non spatially-dependent as we work within the dipole approximation). For this, we consider different orientations between each single molecule and the laser polarization. After orientational averaging, the response of the chiral molecules is obtained at one point of space. The macroscopic response consists in scaling the microscopic response to the local electric field at each point of space to obtain the near-field emission and its propagation to the far field, which will be described in the following sections.

## 2 Molecular response

The time-dependent single-molecule response to the laser field is computed using real-time time-dependent density functional theory (TDDFT) via the Octopus code [1-4]. The calculations have been performed for propylene oxide molecules and considering the local-density approximation [5-7] to account for electronic exchange and correlation effects. In addition, the averaged-density self-interaction correction [8] is employed to describe the long-range behavior of the electron density. Since carbon and oxygen are relatively heavy atoms, their 1s orbitals are barely affected by the laser field, so they were described using pseudo-potentials. The Kohn-Sham orbitals and the electron density were discretized on a uniform real-space grid within a sphere with a radius of 42 a.u. and a spacing of 0.4 a.u. In order to avoid reflections of the electron density at the edges of the sphere, we used a complex absorbing potential with a width of 20 a.u. and a height of -0.2 a.u. The electric field that we employ is elliptically polarized (with an ellipticity of $\varepsilon = 0.05$) and has a Gaussian envelope of 7 fs pulse duration (full-width half maximum), a wavelength of $\lambda = 780$ nm and an amplitude of $E_0 = 0.0414$ a.u.

The time-dependent dipoles were obtained in the molecular frame for 200 molecular orientations with respect to the laser polarization. The orientations were selected considering $N_L = 50$ Lebedev grid points (order 11) in a sphere of $\phi$ and $\theta$ angles, which determines the direction of the main component of the laser field with respect to the molecular frame, and $N_R = 4$ rotations in the $\delta$ coordinate to describe the directions of other orthogonal field components.

From the time-dependent dipoles, we obtain the time-dependent dipolar accelerations. Then, the orientationally averaged dipolar acceleration is obtained as:

$$\vec{a}(t) = \frac{1}{2\pi}\sum_{i=0}^{N_L}\sum_{j=1}^{N_R} \vec{a}_{i,j}(t) w_i \, d\delta \tag{S1}$$

where $w_i$ are the weights of the Lebedev quadrature and $\vec{a}_{i,j}(t)$ is the 3-dimensional, time-dependent dipolar acceleration of a molecule interacting with a laser whose orientation is described by $\phi_i$, $\theta_i$, and $\delta_j$ with respect to the molecular frame.

The modulus to the square of the orientationally averaged dipolar acceleration is proportional to the power of the emitted radiation through the Larmor formula, therefore its Fourier components provide the harmonic spectrum.

## 3 Harmonic spectra

We perform a Fast Fourier Transform (FFT) of the time-dependent emitted power to calculate the harmonics. In figure S1(a), we can observe that the achiral components are composed of odd harmonics, and the emitted electric field is 20 times stronger in the $\rho$ direction than in the $\phi$ direction, because we consider an elliptically polarized electric field. Due to this ellipticity and the chirality of the sample, even harmonics appear in the direction that is orthogonal to the laser's polarization, as shown in figure S1(b). That component carries the enantio-sensitive phase.

It is worth noting that the longitudinal (chiral) component can interfere with the achiral components because the pulse is ultrashort and therefore broadband, so the even and odd harmonics overlap, as shown in figure S1(c). Nevertheless, longer pulses could also be employed as long as they carry a broad spectral range, which can be achieved using chirped laser pulses. Another alternative would be to use two-colour pulses (one frequency to drive the achiral component and the other for the chiral one, resulting in a matching harmonic frequency), so the duration of the pulse can be arbitrary.

In our configuration, the resulting intensity profiles are slightly more enantio-sensitive for q = 5.8 than at the maximum of the chiral spectra, q = 5.897 (see figure S2).

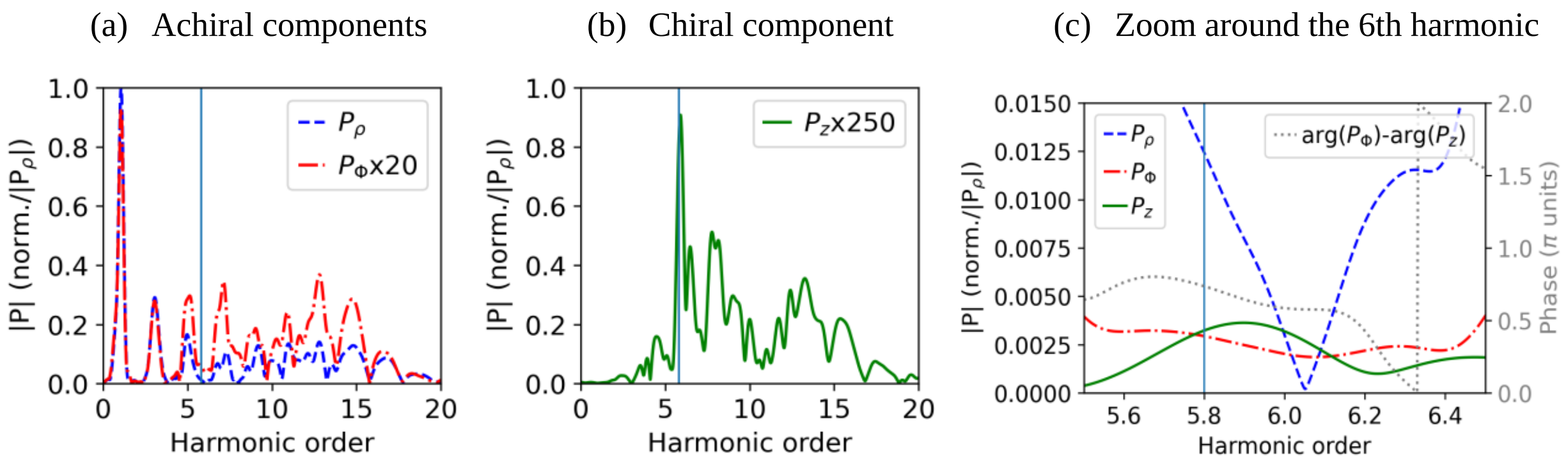

**Fig. S1.** Spectra of the emitted harmonics: (a) achiral polarization components $\rho$ and $\phi$ and (b) chiral polarization component $z$, in which direction the enantio-sensitive HHG appears. (c) In the vicinity of the 6th harmonic, the three components have similar amplitude. The blue line represents where the HHG is more enantio-sensitive ($q = 5.8$).

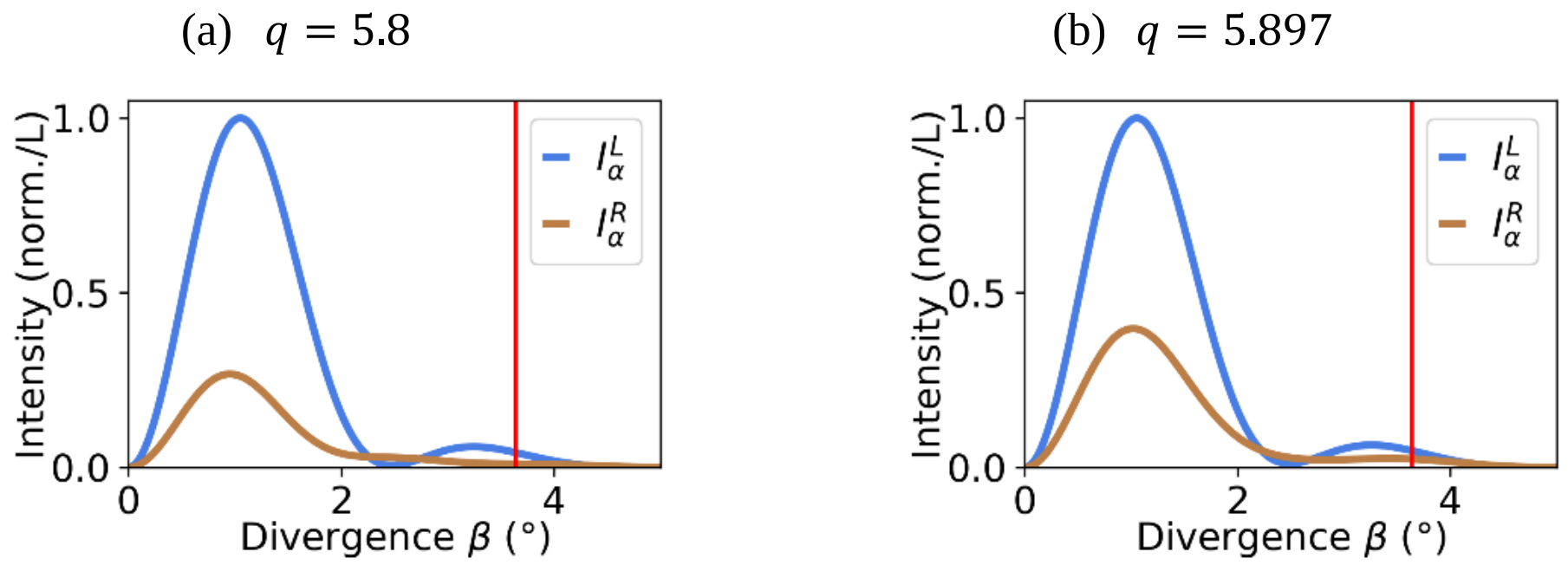

**Fig. S2.** Intensity profile of the azimuthal component emitted by the L- (blue) and R-enantiomers (brown) for (a) $q = 5.8$, where the two emissions differ the most, and (b) $q = 5.897$, corresponding to the maximum of the chiral response. The red line indicates the maximum of the driving IR beam.

Note that the signal at q = 5.8 is about three orders of magnitude lower in amplitude than the linear response, for both the chiral and achiral components. This is the typical order of magnitude in conversion efficiency in HHG experiments driven in gases such as Argon [9], so its measurement is within the standard experimental capabilities. Table S1 gives the amplitude values at q = 5.8 that were used in our calculations.

| | Absolute values ($\times 10^{-12}$ a.u.) | Relative values ($\lvert P\rvert/\max\lvert P_\rho\rvert$) |
|---|---|---|
| $\lvert P_{\rho,o}\rvert$ | 2.0056 | 0.01246 |
| $\lvert P_{\phi,o}\rvert$ | 0.4736 | 0.00294 |
| $\lvert P_{z,o}\rvert$ | 0.5200 | 0.00323 |

**Table S1.** Spectral power at $q = 5.8$ obtained from the TDDFT calculations. The relative amplitudes correspond to the absolute value normalized with respect to the linear response (first harmonics).

## 4 Vector beam definition

The ellipticity of the driving beam is achieved by combining a radially polarized vector beam and an azimuthally polarized one with a phase delay of $\pi/2$. We define the amplitude ratio between the two as $\varepsilon_\phi$. Both of the vector beams are defined as a superposition of two circularly-polarized vortex beams with opposite helicity and topological charge $\ell$, and radial mode p:

$$\begin{cases} \vec{E}_\rho(\rho,\phi,z) = LG_{\ell,p}(\rho,\phi,z)(\hat{e}_x + i\hat{e}_y) + LG_{-\ell,p}(\rho,\phi,z)(\hat{e}_x - i\hat{e}_y) \\ \vec{E}_\phi(\rho,\phi,z) = i\varepsilon_\phi[LG_{\ell,p}(\rho,\phi,z)(\hat{e}_x + i\hat{e}_y) - LG_{-\ell,p}(\rho,\phi,z)(\hat{e}_x - i\hat{e}_y)]. \end{cases} \tag{S2}$$

In this work, we consider radially and azimuthally polarized vector beams (i.e. with a Poincaré index equal to 1, so $\ell = 1$) at the focal plane ($z = 0$) and with no radial nodes ($p = 0$). Therefore, in polar unitary vectors (see Fig. 1 in the main text), the expressions of the vector beams become independent of $\phi$:

$$\begin{cases} \vec{E}_\rho(\rho) = 2R_{\ell=1,p=0}(\rho)\hat{e}_\rho \\ \vec{E}_\phi(\rho) = 2i\varepsilon_\phi R_{\ell=1,p=0}(\rho)\hat{e}_\phi, \end{cases} \tag{S3}$$

where $R_{\ell,p}(\rho)$ is the $\rho$-dependent part of the Laguerre-Gaussian mode formula [10,11] such that $LG_{\ell,p}(\rho,\phi) = R_{\ell,p}(\rho)e^{i\ell\phi}$ and:

$$R_{1,0}(\rho) = E_0\frac{\sqrt{2}\rho}{w_0}e^{-\frac{\rho^2}{w_0^2}}, \tag{S4}$$

$E_0$ being an amplitude constant.

Upon tight focusing, a longitudinal component appears [12]:

$$\vec{E}_z(\rho,\phi) = \frac{i}{k}[\vec{\nabla}_T \cdot \vec{E}_T(\rho,\phi)]\,\hat{e}_z, \tag{S5}$$

where the T indicates transverse.

For our particular case of $z = 0$, $\ell = 1$, $p = 0$, only the radially polarized beam exhibits a longitudinal component:

$$\vec{E}_z(\rho) = 2\frac{i}{k}\frac{\sqrt{2}}{w_0}E_0\left(2 - \frac{2\rho^2}{w_0^2}\right)e^{-\frac{\rho^2}{w_0^2}}\hat{e}_z. \tag{S6}$$

So $E_z$ has a radial node at $\rho = w_0$, hence the second ring pictured in Fig. 1(c) in the main text, where the polarization is reversed. For a peak intensity of $E_\rho$ of $|E_\rho|^2 = 5.1 \times 10^{13}$ W/cm$^2$, the intensity maximum of $E_z$ reaches $2.26 \times 10^{12}$ W/cm$^2$. As a comparison, for the same $E_0$ and $w_0$, the longitudinal component of a Gaussian beam reaches an intensity maximum of $7.65 \times 10^{10}$ W/cm$^2$, i.e. 29 times weaker. Therefore, vector beams present a stronger longitudinal component compared to Gaussian beams [13], so they are a better candidate for our approach.

The robustness of vector beams upon propagation has been studied as the conservation of the polarization inhomogeneity of vector beams, which is more resistant upon propagation through complex media, compared to other physical quantities (amplitude, phase, ellipticity) [14]. In addition, the axial symmetry of vector beams and the co-propagation of the harmonic emissions and driving field makes the setup itself more robust, compared to other chirality studies which require non-collinear configurations [15] or tightly focused Gaussian beams [16], both leading to separate propagation directions of the harmonic emission.

## 5 Interferences in the near field

The interaction with the macroscopic medium is simulated by considering a thin gas jet, but our method can also be applied to a liquid thin sheet. A thin medium would be the best choice to maximize the interaction with the laser's longitudinal component, which is the strongest at the focus. In a thicker medium, the contribution of the achiral signal would be stronger than the chiral one at the interaction planes that are further from the focus. Nevertheless, the field is less intense in those regions, and a stronger achiral signal can be compensated with a lower waist (i.e. more tightly focused beam, which exhibits a stronger longitudinal component). Experimentally, very thin liquid sheets, of 1.5-micron thickness, can be realized [17]. Under these conditions, the maximum of the longitudinal component only decreases by 0.03% at the edges of the thin sheet, while the Gouy phase only changes by 0.025 rad. For these reasons, our calculations are performed considering a 2D sample.

To obtain the macroscopic harmonic emission in the near field, we take into account that the enantio-sensitive HHG reaches its maximum for a harmonic that is still in the perturbative region of the harmonic spectra. This allows a perturbative scaling of the harmonics with the local intensity. Thus, we apply to the dipolar acceleration the perturbative scaling of the nonlinear induced polarization $P_i = \epsilon_0 (\sum_j \chi_{ij}^{(1)} E_j + \sum_{jk} \chi_{ijk}^{(2)} E_j E_k + \sum_{ijkl} \chi_{ijkl}^{(3)} E_j E_k E_l \ldots)$, where $\epsilon_0$ is the vacuum permittivity, $\chi^{(m)}$ are the mth order susceptibilities and the subscripts indicate the induced polarization direction (i) and the electric field's polarization components (j,k,l)$\in \{\rho, \phi, z\}$. Note that this procedure can be extended to the non-perturbative part of the spectrum by realizing a non-perturbative scaling. In our case, the electric field for the harmonic q scales perturbatively with the local electric field in the tilted frame, that we call laser frame $(\rho, \varphi, \zeta)$, as:

$$\begin{cases} E_\rho^{(q)}(\rho) &= \left|\dfrac{E_\rho(\rho)}{E_{\rho,o}}\right|^q P_{\rho,o}^{(q)} \, \mathrm{sign}[E_\rho(\rho)] \\ E_\varphi^{(q)}(\rho) &= \left|\dfrac{E_\rho(\rho)}{E_{\rho,o}}\right|^q \dfrac{|\varepsilon_\varphi(\rho)|}{\varepsilon_{\phi,o}} P_{\phi,o}^{(q)} \, \mathrm{sign}[E_\phi(\rho)] \\ E_\zeta^{(q)}(\rho) &= \left|\dfrac{E_\rho(\rho)}{E_{\rho,o}}\right|^{q+2} \dfrac{|\varepsilon_\varphi(\rho)|}{\varepsilon_{\phi,o}} P_{z,o}^{(q)} \, \mathrm{sign}[\varepsilon_\phi], \end{cases} \tag{S7}$$

with $E_{\rho,o}$ the laser amplitude used for TDDFT calculations and $P_{\{\rho,\phi,z\},o}^{(q)}$ the resulting electric field for the qth harmonic. The harmonic field does not depend on $\phi$ due to the axial symmetry of the vector beam (see equation S3). The response is also scaled to the local ellipticity $\varepsilon_\varphi(\rho)$ with respect to that used in the TDDFT calculations $\varepsilon_{\phi,o} > 0$, taking into account that it only changes sign if the transversal ellipticity, $\varepsilon_\phi$, does. Note that, while the $\rho$ and $\varphi$-components follow the usual scaling for odd-order (achiral) harmonics, the $\zeta$-component scales as the $(q+2)$th power because it results from the absorption of $q+1$ photons from the main laser component and 1 photon from the elliptical component [15] and $\left|\frac{E_\rho(\rho)}{E_{\rho,o}}\right|^{q+1} \left|\frac{E_\varphi(\rho)}{E_{\phi,o}}\right| = \left|\frac{E_\rho(\rho)}{E_{\rho,o}}\right|^{q+2} \frac{|\varepsilon_\varphi(\rho)|}{\varepsilon_{\phi,o}}$. Importantly, $E_{z,o}^{(q)}$ changes sign for opposite enantiomers.

In the laboratory frame, the tilt angle satisfies $\cos[\gamma(\rho)] = \frac{\varepsilon_\phi}{|\varepsilon_\varphi(\rho)|}$ and $\sin[\gamma(\rho)] = \frac{\varepsilon_z(\rho)}{|\varepsilon_\varphi(\rho)|}$, where $|\varepsilon_\varphi(\rho)| = \sqrt{\varepsilon_\phi^2 + \varepsilon_z^2(\rho)}$. Thus, the induced polarization in the laboratory frame is:

$$\begin{cases} E_\phi^{(q)}(\rho) = \cos[\gamma(\rho)] E_\varphi^{(q)}(\rho) - \sin[\gamma(\rho)] E_\zeta^{(q)}(\rho) \\ E_z^{(q)}(\rho) = \sin[\gamma(\rho)] E_\varphi^{(q)}(\rho) + \cos[\gamma(\rho)] E_\zeta^{(q)}(\rho) \end{cases} \tag{S8}$$

Since $E_\zeta^{(q)}$ has opposite phase for opposite enantiomers, equation S8 describes a harmonic field that is different for each enantiomer.

Note that our approach could be applied to any molecular species by tuning the laser. The amplitude and phase of the response depend on the nonlinear susceptibilities of the medium (molecule-specific) and the driving field. If one chooses a different molecule or field frequency (i.e. if a different nonlinear susceptibility characterizes the response) the interference between the achiral and chiral harmonic components can be preserved by adjusting the driving field (see equation S8): a larger tilt angle favors the projection of the chiral component over the achiral one. This tilt angle is controlled either via the focusing conditions (stronger longitudinal component leads to higher tilt), or the ellipticity (larger ellipticity reduces the tilt). In addition, higher field intensities favor the chiral component over the achiral one (see equation S7). Thus, any of these three quantities can be used as a control knob to maximize the enantio-sensitivity by balancing the achiral and chiral contributions.

On the other hand, the chirality of the sample ensures that the phase of the chiral component is opposite for opposite enantiomers regardless of the molecular species and the driving field (as it this is protected by the symmetry of the sample and the setup). Thus, even though the exact phase difference between the chiral and achiral components depends on the selected molecular and harmonic order, it will always be opposite for opposite enantiomers, which guarantees that the emitted intensity profile is enantio-sensitive once the amplitudes of the chiral and achiral components are of similar magnitude.

## 6 Propagation to the far field

To obtain the nonlinear emission at the detection plane (far field), we apply the Fraunhofer diffraction equation, valid for an observer at a distance $z_{FF} >> \frac{q\rho^2}{\lambda_0} \approx 2$ mm from the target. The azimuthal and radial far-field coordinates are denoted as α and β, respectively. The two field polarization components in the far-field are denoted as $E_\alpha^{FF}(\alpha,\beta)$ and $E_\beta^{FF}(\alpha,\beta)$ and they are obtained as:

$$E_{\alpha/\beta}^{FF}(\alpha,\beta) \propto \int_0^\infty \int_0^{2\pi} A_{\alpha/\beta}^{NF}(\alpha,\rho,\phi) e^{-i\frac{2\pi}{\lambda_0} q \tan\beta \cos(\alpha-\phi)\rho} \rho \, d\rho \, d\phi \quad \text{(S9)}$$

with q being the harmonic number, $\lambda_0 = 780$ nm the fundamental wavelength, and $A_{\alpha/\beta}^{NF}$ the quantity containing the near-field components (Eq. S7 and S8) as:

$$\begin{cases} A_\beta^{NF}(\alpha,\rho,\phi) = \cos(\phi-\alpha)E_\rho^{(q)}(\rho) + \sin(\phi-\alpha)E_\phi^{(q)}(\rho) \\ A_\alpha^{NF}(\alpha,\rho,\phi) = \sin(\alpha-\phi)E_\rho^{(q)}(\rho) + \cos(\phi-\alpha)E_\phi^{(q)}(\rho). \end{cases} \quad \text{(S10)}$$

We apply the same procedure for both left and right enantiomers.

## 7 Concentration of each enantiomer

In complement to Fig. 3(e) in the main text, the figure S3 depicts the evolution of the radius and the value of the intensity maximum for different concentrations of left-handed enantiomer $C_L$ (the rest of the solution is composed of right-handed enantiomer $C_R = 1 - C_L$).

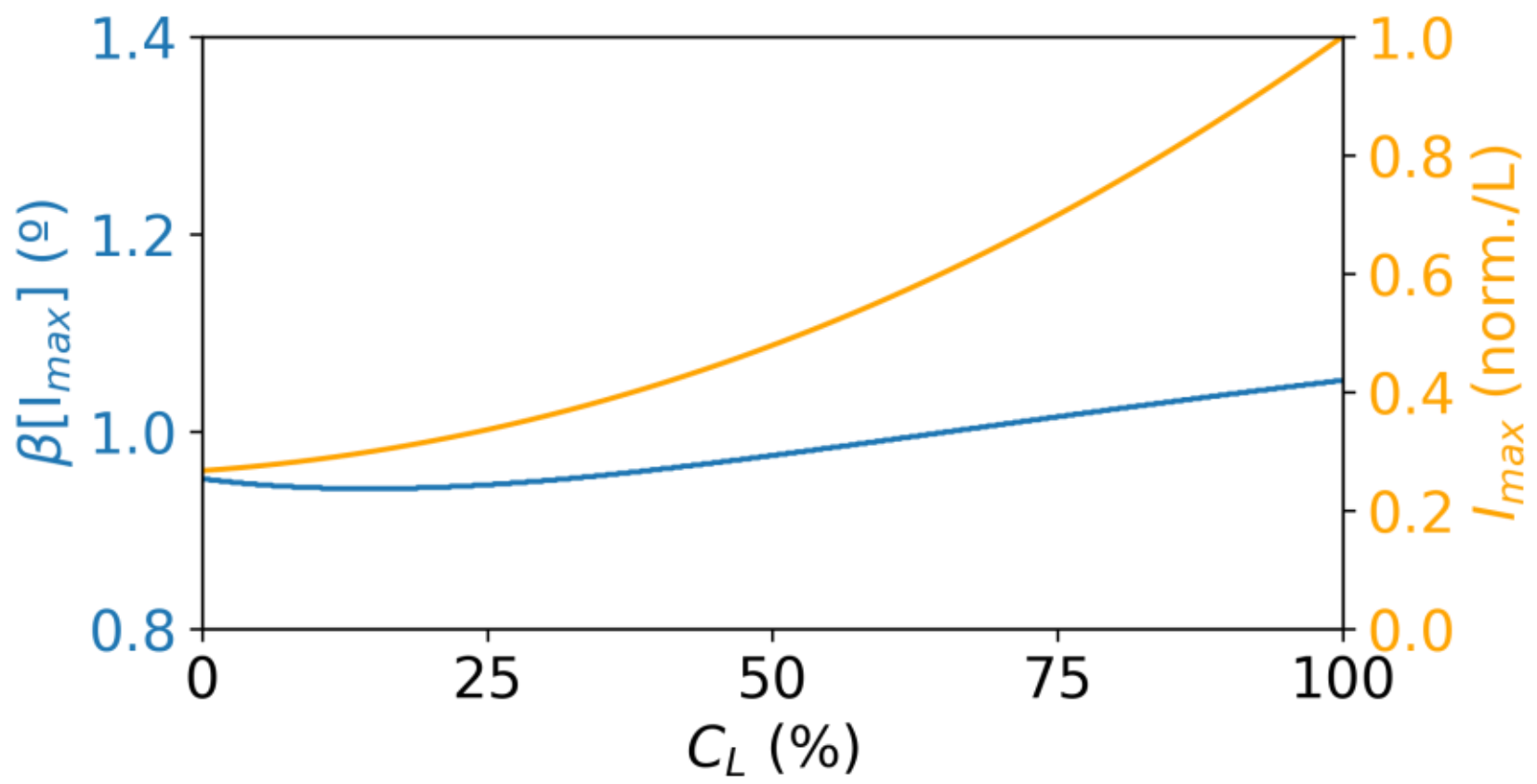

**Fig. S3.** Radius of the intensity maximum (blue) and value reached for this maximum (orange) of the $q$th harmonic for different proportions of R-enantiomers $C_R = 1 - C_L$.

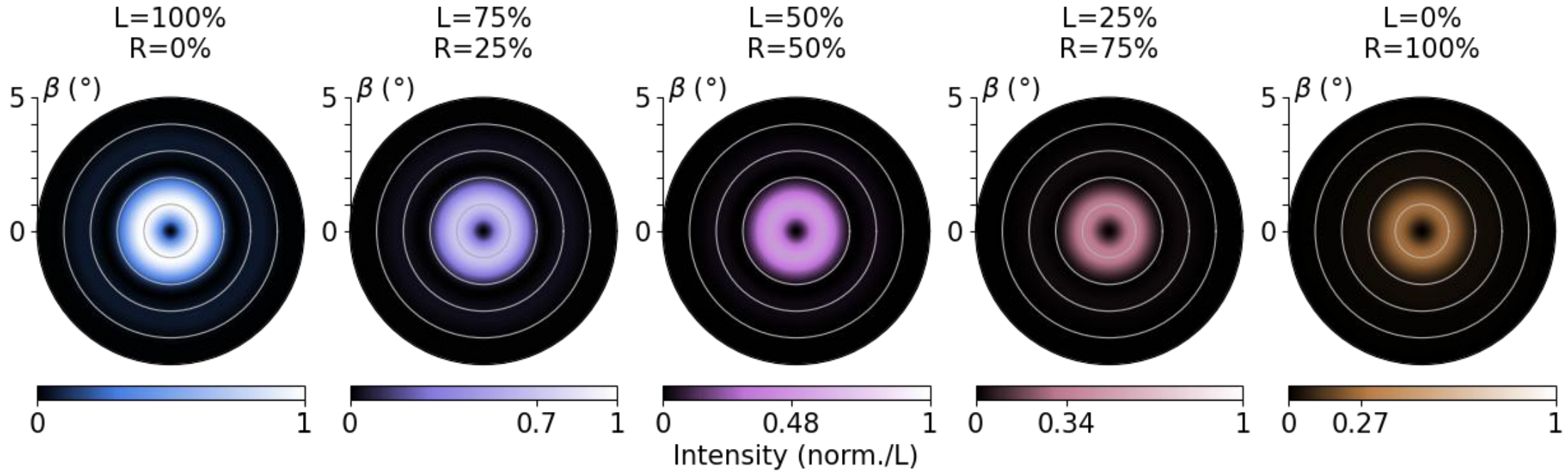

**Fig. S4.** Intensity profiles the $q$th harmonic in the far field for different concentration of each enantiomer. All profiles have been normalized with respect to the left figure, shown for pure L-enantiomer. The resulting maxima of intensity are written in each colorbar, and they correspond to the yellow curve in figure S3.

To gain further insight, we also show, in figure S4, how changing the concentration of left-handed or right-handed molecules changes the intensity profile in the far field. We can thus distinguish the proportion of each enantiomer by the peak intensity and shape of the emission.

## 8 Ellipticity of the harmonic emission in the far field

The arrows in the figure 3(d) of the main text were plotted using the ellipticity calculated with the Stokes parameters. The full ellipticity distribution can be appreciated in figure S5.

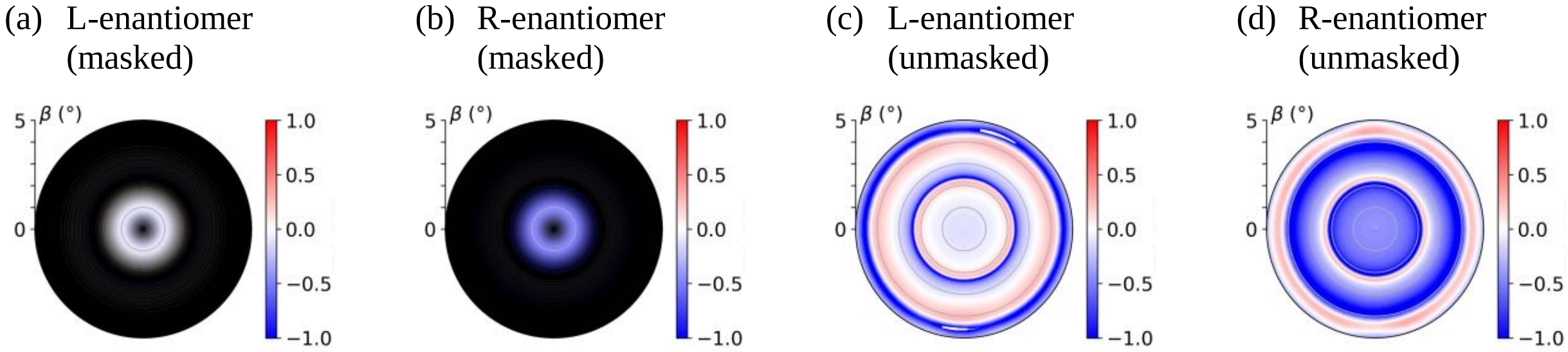

**Fig. S5.** Ellipticity extracted from the Stokes parameters for the (a,c) L- and (b,d) R-enantiomers. A transparency mask of the total intensity (radial and azimuthal polarizations) has been applied to (a) and (b) to highlight the most important regions of the figures.

**References.**